\documentclass[10pt]{amsart}
\usepackage{amscd}
\newtheorem{thm}{Theorem}

\newtheorem{lem}[thm]{Lemma}
\newtheorem{cor}[thm]{Corollary}

\theoremstyle{remark}
\newtheorem{remark}[thm]{Remark}
\theoremstyle{definition}
\newtheorem{dfn}[thm]{Definition}
\newtheorem{example}[thm]{Example}

\newcommand{\FF}{\mathcal{ F}}
\newcommand{\OO}{\mathcal{ O}}
\newcommand{\UU}{\mathcal{ U}}
\newcommand{\T}{\mathbb{ T}}
\newcommand{\R}{\mathbb{ R}}

\newcommand{\g}{\mathfrak{ g}}

\DeclareMathOperator{\rank}{rank}

\DeclareMathOperator{\sgrad}{sgrad}
\DeclareMathOperator{\reg}{reg}
\DeclareMathOperator{\ddim}{ddim}
\DeclareMathOperator{\dind}{dind}

\begin{document}

\title{On the integrability of geodesic flows of submersion metrics
\footnote{Journal Ref.: Letters in Mathematical Physics {\bf 61}:
29-39, 2002.}}

\author{Bo\v zidar~Jovanovi\' c}

\address{Mathematical Institute SANU,
Kneza Mihaila 35, Belgrade, Serbia, Yugoslavia}
\email{bozaj@mi.sanu.ac.yu}


\abstract
Suppose we are given a compact Riemannian manifold $(Q,g)$ with a completely
integrable geodesic flow. Let $G$ be a compact connected Lie group
acting freely on $Q$ by isometries.
The natural question arises:
will the geodesic flow on $Q/G$ equipped with the submersion metric
be integrable?
Under one natural assumption, we prove
that the answer is affirmative.
New examples of manifolds with completely integrable geodesic flows
are obtained.
\endabstract

\maketitle

\section{Introduction}

Let $T^*Q$ be the cotangent bundle of a compact connected
$n$--dimensi\-onal Riemannian manifold
$(Q,g)$ with the natural symplectic structure.
The geodesic flow on $Q$ is described by the Hamiltonian equations
on $T^*Q$
\begin{equation}
\dot x=\sgrad H(x),
\label{0.1}
\end{equation}
where the Hamiltonian function is $H(p,q)=\frac12g^{-1}_q(p,p)$, $p\in T^*_q Q$.
The geodesic flow (\ref{0.1}) is {\it completely integrable}
if  there are
$n$ Poisson-commuting smooth integrals $f_1,\dots,f_n$ whose
differentials are independent in an open dense set of $T^*Q$.
Then by Liouville's theorem $T^*Q$ is
foliated by invariant Lagrangian tori in the open dense set.
This situation is very exceptional.
Furthermore, if additional conditions are placed
on the integrals (such as real analyticity) then there are
serious topological obstructions to the integrability
\cite{K, T, P1}.
Examples of manifolds admitting  completely integrable
geodesic flows are given in \cite{Thimm, Br, TF, PS, Baz, BT, BT1, BJ1, BJ}.
A more detailed list of references can be found in  \cite{BJ1, BJ}.

Suppose we are  given a compact Riemannian manifold $(Q,g)$ with completely
integrable geodesic flow. Let $G$ be a compact connected Lie group
acting freely on $Q$ by isometries.
The problem we are interested in is as follows:
Will the geodesic flow on $Q/G$, equipped with the submersion metric,
be integrable?

Paternain and Spatzier proved that if the manifold $Q$ has
geodesic flow intergrable
by means of $S^1$--invariant integrals and if $N$ is a
surface of revolution, then the submersion
geodesic flow on  $Q\times_{S^1} N=(Q\times N)/S^1$
will be completely integrable  \cite{PS}.
Combining submersions
and Thimm's method (see \cite{Thimm}),
Paternain and Spatzier \cite{PS} and Bazaikin \cite{Baz}
proved integrability of geodesic flows on
certain interesting  bi-quotients of Lie groups
(in fact, the author's motivation
in writing this letter was to explain these submersion examples
in the framework of a general construction).

In this letter we use the Mishchenko--Fomenko--Nekhoroshev
theorem on
non-commutative integration of Hamiltonian systems
(section 2).
This allows us to take a new approach to the proposed problem
and to prove that the answer,
under one natural assumption, is affirmative.
In section 3, the simple proof is given
if original integrals are $G$--invariant functions
(which forces $G$ to be
abelian) and in section 4, as the main result, the proof
is given for the arbitrary $G$--actions.
As an illustration,
we show the integrability of geodesic flows on a
class of bi-quotients of Lie groups (section 5).

\section{Local structure of integrable systems}

We shall briefly recall the concept of non-commutative integrability introduced by Mishchenko and Fomenko in \cite{MF}.
Let $M$ be a $2n$--dimensional symplectic manifold.
Let $(\FF,\{\cdot,\cdot\})$ be a Poisson subalgebra
of  $(C^\infty(M),\{\cdot,\cdot\})$.
Let $F_x$ be the subspace of $T^*_xM$ generated by $df(x)$, $f\in \FF$.
Suppose that  on an open dense set of
$M$ we have $\dim F_x=l$ and $\dim \ker\{\cdot,\cdot\}\vert_{F_x}=r$.
Denote the numbers $l$ and $r$ by $\ddim\FF$ and $\dind\FF$,
respectively.
The algebra $\FF$ is {\it complete} if
the following condition is satisfied:
$$
\ddim{\FF}+\dind{\FF}=\dim M.
$$

\begin{remark}
Let $\FF$ be a complete algebra. Set
\begin{eqnarray*}
&&W_x=\{\sgrad f(x),\; f\in {\FF}\}, \\
&&D_x=\{ \xi \in T_x M,\; df(x)(\xi)=0, f\in\FF\}.
\end{eqnarray*}

The condition
$\dim F_x+\dim \ker \{\cdot,\cdot\}_{F_x}=\dim M$ is
equivalent to the coisotropy of $W_x$ and isotropy of $D_x$
in the symplectic linear space $T_xM$.
In this case we say that $\FF$ is {\it complete at} $x$.
\end{remark}

\smallskip

The Hamiltonian system $\dot x=\sgrad H(x)$ on
a symplectic manifold $M$
is called {\it completely integrable (in the non-commutative sense)} if it
possesses a complete
algebra of first integrals $\FF$.
Then  each connected compact component of a regular level set of the
functions $f_1,\dots,f_l\in\mathcal F$
is an $r$--dimensional invariant isotropic torus $\mathbb{T}^r$
(see \cite{MF, N, TF}).
In a neighborhood of $\T^r$ there are {\it generalized action-angle
variables} $p,q,I,\varphi\mod2\pi$, defined in a toroidal domain
$\OO=\T^r\{\varphi\}\times B_\sigma\{I,p,q\}$,
$$
B_\sigma=\{(I_1,\dots,I_r,p_1,\dots,p_k,q_1,\dots,q_k)\in\R^l,\;
\sum_{i=1}^r I_i^2 + \sum_{j=1}^k q_j^2+p_j^2 \le \sigma^2\}
$$
such that the symplectic form becomes
$$
\omega=\sum_{i=1}^r dI_i\wedge d\varphi_i+\sum_{i=1}^k dp_i\wedge dq_i,
$$
and the Hamiltonian function depends only on $I_1,\dots,I_r$.
The Hamiltonian equations take the following form in action-angle
coordinates:
\begin{equation}
\dot \varphi_1=\omega_1(I)=\frac{\partial H}{\partial I_1},
\dots, \dot \varphi_r=\omega_r(I)=\frac{\partial H}{\partial I_r},
\quad \dot I=\dot p=\dot q=0.
\label{2.1}
\end{equation}

\begin{dfn}  \cite{Bo} The Hamiltonian system (\ref{2.1})
defined in the toroidal domain $\OO=\T^r\times B_\sigma$ is {\it $\T^r$--dense}
if the set of points $(I_0,p_0,q_0)\in B_\sigma$ for which the
trajectories of (\ref{2.1}) are dense on the torus $\{I=I_0,p=p_0,q=q_0\}$
is everywhere dense in the ball $B_{\sigma}$.
\end{dfn}

The frequencies $\omega_1(I),\dots,\omega_r(I)$ corresponding
to the dense trajectories (\ref{2.1}) are rationally independent.
For example,  a non-degenerate system
($\det\left(\frac{\partial \omega}{\partial I}\right)\ne 0$ on an
open dense set of $\OO$) is $\T^r$--dense.

Any smooth first integral of a $\T^r$--dense system
is a function of the variables $I,q,p$ only.

It follows from theorem 8 in Bogoyavlenskij \cite{Bo}
that for an arbitrary differentiable Hamiltonian function
$H(I_1,\dots,I_r)$ defined on a toroidal domain
$\OO=\T^r\{\varphi\}\times B_\sigma\{I,p,q\}$,
there exists a family of balls $B_\tau\subset B_\sigma$ such that
the union $\cup_\tau B_\tau$ is dense in $B_\sigma$ and the following
properties hold.
In the toroidal domain $\OO_\tau=\T^r\times B_\tau$
there exists a canonical transformation:
$\{I,p,q,\varphi\}\to\{I^\tau,p^\tau,q,^\tau,\varphi^\tau\}$,
that transforms the system to the form
$$
\dot\varphi_1^\tau=\omega_1^\tau=\frac{\partial H}{\partial I_1^\tau},
\dots,\dot\varphi_{\tilde r(\tau)}^\tau=\omega_{\tilde r}^\tau=
\frac{\partial H}{\partial I_{\tilde r}^\tau},
$$
$$
\dot\varphi_{\tilde r+1}^\tau=0,\dots,\dot\varphi_r^\tau=0, \quad
\dot I^\tau=\dot p^\tau=\dot q^\tau=0,
$$
$H=H(I_1^\tau,\dots,I_{\tilde r})$, $\tilde r=\tilde r(\tau)\le r$.
The system is $\T^{\tilde r}$--dense in $\OO_\tau$
(regarded as the product $\T^{\tilde r}\times
(\T^{r-\tilde r}\times B_{\tau})$).
Moreover, if $H(I_1,\dots,I_r)$ is analytic and the maximal dimension
of the closures of trajectories is equal to $\tilde r$ then such canonical
transformation exists globally and the system is $\T^{\tilde r}$--dense in $\OO$.

We turn back to geodesic flows.
Let $(Q,g)$ be a compact Riemannian manifold and let
$H(p,q)=\frac12 g_q^{-1}(p,p)$.
Suppose that the geodesic flow is completely integrable
in the non-commutative sense by means of a complete
algebra $\FF$ of first integrals.
Since all iso-energy levels $(T^*Q)_h=H^{-1}(h)$ are compact,
the phase space $T^*Q$ is
foliated by invariant $\dind\FF$--dimensional isotropic tori
in an open dense set
which we shall denote by $\reg T^*Q$.
Although the dynamics on $\reg T^*Q$ are well understood,
 examples by Bolsinov and Taimanov \cite{BT, BT1}
show that an integrable geodesic flow may possess complicated dynamics
on the singular set $T^*Q\setminus \reg T^*Q$.

If the algebra of integrals $\FF=\{f_1,\dots,f_n\}$ is commutative
($\ddim\FF=\dind\FF=n$) then we have usual
Liouville integrability.
But if the geodesic flow (\ref{0.1})
is integrable in the non-commutative sense, then
it is also integrable in the commutative sense, i.e.,
there exist $C^\infty$--smooth commuting integrals $g_1,\dots,g_n$
that are independent on an open dense set of $T^*Q$.
The $r$--dimensional invariant isotropic tori $\T^r$
can be organized into larger, $n$--dimensional Lagrangian tori $\T^n$
which  are the level sets of the commutative algebra of integrals
$\{g_1,\dots,g_n\}$
\cite{BJ}.
Furthermore, if $\FF=\{f_1,\dots,f_l\}$
is a finite dimensional Lie algebra ($\{f_i,f_j\}=c^k_{ij}f_k$),
then the commuting integrals can be taken as polynomials in $f_1,\dots,f_l$
(see Brailov \cite{Br}).

\section{Submersions}

Let $G$ be a compact connected Lie group
with a free Hamiltonian action on the symplectic manifold $(M,\omega)$.
Let
$$
\Phi: M \to \g^*
$$
be the corresponding equivariant moment map
($\g^*$ is the dual space of the Lie algebra $\g=T_eG$).
Let $G_\eta$ be the  coadjoint action isotropy group of
$\eta\in\Phi(M)\subset \g^*$.
By $(M_\eta,\omega_\eta)$ we denote the reduced symplectic space
$$
M_\eta=\Phi^{-1}(\eta)/G_\eta,
$$
$$
\omega_\eta(d\pi(\xi_1),d\pi(\xi_2))=\omega(\xi_1,\xi_2), \;
\xi_1,\xi_2\in T_x\Phi^{-1}(\eta),
$$
where $\pi: \Phi^{-1}(\eta)\to M_\eta$ is the natural projection.
For a $G$--invariant function $f$ on $M$,
let $f_\eta$ be the induced function on $M_\eta$.

We have the following simple general observation.

\begin{lem}
Suppose
a connected compact Lie group $G$ acts effectively on
a symplectic manifold $(M,\omega)$ with moment map $\Phi$.
Let $H$ be a $G$--invariant Hamiltonian function.
If the Hamiltonian system
$\dot x=\sgrad H(x)$ is completely integrable by means of
$G$--invariant first integrals, then $G$ is a torus.
Moreover,
if the action is free,
the reduced Hamiltonian system on $M_\eta$ is completely
integrable for a generic value $\eta$ of the moment map.
\end{lem}

\begin{proof}
Let $\FF$ be a complete $G$--invariant algebra of integrals
and let $D_x$ and $W_x$ be the subspaces of $T_xM$
defined in remark 1.
Since all functions from $\FF$ are $G$--invariant we have that
$T_x(G\cdot x)\subset D_x$ is isotropic,
for generic $x\in M$.
The Hamiltonian vector fields of the functions
$\phi_a(x)=\langle \Phi(x), a \rangle$, $a\in\g$
generate
one-parameter groups of symplectomorphisms.
Therefore
$\omega_x(\sgrad \phi_{a}(x),\sgrad \phi_{b}(x))$
vanishes
in an open dense set of $M$. Thus $\phi_{[a,b]}=
\{\phi_{a},\phi_{b}\} = 0$ for every $a,b \in \g$.
Since the action is effective, we get that $G$ is a torus.

Let the torus $G$ act freely on $M$.
Take $x\in M$ such that $\FF$ is complete at $x$.
Since all functions from $\FF$ are $G$--invariant,
the coisotropic space $W_x=\{\sgrad f(x), \; f\in \FF\}$
belongs to  $T_x \Phi^{-1}(\eta)$, $\eta=\Phi(x)$.
Let $\pi: \Phi^{-1}(\eta)\to M_\eta$ be the natural projection.
From the definition of the reduced symplectic structure,
one can easily see that the space
generated by $\sgrad f_\eta(\pi(x))$, $f\in \FF$
is also coisotropic, i.e., the induced algebra $\FF_\eta=\{f_\eta, f\in\FF\}$
is complete at $\pi(x)$.
\end{proof}

Now let a compact connected Lie group $G$ act on a
Riemannian manifold $(Q,g)$ by isometries.
Suppose that $G$ acts freely on $Q$
and endow $Q/G$ with the submersion metric. Let
$\mathcal G_x+\mathcal H_x=T_x Q$
be the orthogonal decomposition of $T_xQ$, where $\mathcal G_x$
is tangent space to the fiber
$G \cdot x$. By definition, the submersion metric $g_{sub}$ is given by
$$
\langle \xi_1,\xi_2 \rangle_{\rho(x)} =
\langle \bar \xi_1,\bar \xi_2 \rangle_x,
\quad \xi_i \in T_{\rho(x)}(Q/G), \; \bar \xi_i \in \mathcal H_x, \;
\xi_i=d\rho(\bar \xi_i),
$$
where $\rho: Q \to Q/G$ is the canonical projection. The vectors in
$\mathcal G_x$ and $\mathcal H_x$ are called
{\it vertical} and  {\it horizontal} respectively.

Let $\Phi: T^*Q\to \g^*$ be the moment map of the natural
Hamiltonian $G$--action on $T^*Q$.
It is well known that the reduced symplectic
space
$$
(T^*Q)_0=\Phi^{-1}(0)/G
$$
is symplectomorphic to $T^*(Q/G)$.
Moreover, if $H$ is the Hamiltonian function of the geodesic
flow on $Q$ then $H_0$ will be the Hamiltonian of the geodesic
flow for the submersion metric.
If we identify $T^*Q$ and $TQ$ by the metric $g$, then
$\Phi^{-1}(0)$ will be
the set of all horizontal vectors $\mathcal H$ (see \cite{PS}).

Therefore, by lemma 3,
if the geodesic flow of $(Q,g)$
is completely integrable by
means of a $G$--invariant algebra of integrals $\FF$
then $G$ is commutative.
Moreover we have integrability
of the reduced system on $(T^*Q)_\eta$, for generic $\eta\in\Phi(T^*Q)$.
Since $G$ is commutative $(T^*Q)_\eta$ is diffeomorphic to
$T^*(Q/G)$, but for $\eta\ne 0$
the symplectic form possesses the additional "magnetic term".
The reduced flow is no longer the inertial motion of a particle
on $Q/G$ (i.e., geodesic flow), but the motion of a particle
under an additional magnetic force.
Specifically, if $\reg T^*Q$ intersects the space of horizontal vectors
$\mathcal H\cong \Phi^{-1}(0)$ in a dense set then
the geodesic flow of $(Q/G,g_{sub})$
will be completely integrable.

We can push the last observation further.

\section{The main theorem}

From lemma 3, the condition on a $G$--invariant algebra $\FF$
on $(M,\omega)$ to be complete forces $G$ to be abelian, which is too
restrictive.
However, it can  happen that $\FF$
is not complete on $(M,\omega)$
but ${\FF}_\eta$ is a complete algebra on the reduced space
$(M_\eta,\omega_\eta)$.

\begin{thm}
Let a compact connected Lie group $G$ act freely by
isometries on the compact Riemannian manifold $(Q,g)$.
Suppose that the geodesic flow is completely integrable.
If $\reg T^*Q$ intersects the space of horizontal vectors
$\mathcal H\cong \Phi^{-1}(0)$ in a dense set then
the geodesic flow on $Q/G$ endowed with the submersion metric $g_{sub}$
is also completely integrable.
\end{thm}

\begin{proof}
Take some toroidal domain
$\OO=\T^r\{\varphi\}\times B_\sigma\{I,p,q\} \subset \reg T^*Q$
($\OO_1=\OO\cap \Phi^{-1}(0) \ne \emptyset$)
such that the geodesic flow is $\T^r$--dense in $\OO$.
Consider the $G$--invariant sets
$$
\UU_1=\UU\cap \Phi^{-1}(0),
\quad \UU=G\cdot \OO=\{g\cdot x,\; x\in \OO,\; g\in G\}.
$$

Note that the functions $\phi_a(x)=\langle \Phi(x), a\rangle$, $a\in \g$
are first integrals of the geodesic flow
and so do not depend on $\varphi$ in $\OO$.
In other words, the action of $G$ preserves the foliation
by  $r$--dimensional invariant isotropic tori of the sets
$\UU$ and $\UU_1$ as well.

The foliation $\mathcal T$ of $\UU_1$ by tori induces a foliation
$\mathcal L$ of $\UU_1$ by
$(\dim G+r-s)$--dimensional
compact $G$--invariant submanifolds, with tangent spaces
of the form
$$
T_x\mathcal L=T_x(G\cdot x)+T_x(\mathcal T)\subset T_x \Phi^{-1}(0),
$$
where
$s=\dim S_x$,
\begin{equation}
S_x=T_x(G\cdot x) \cap T_x(\mathcal T),
\;
x\in \UU_1
\label{*}
\end{equation}
(for $x\in \OO_1$ we have that
$T_x\mathcal L=\{\sgrad \phi_a(x), \; \sgrad I_i(x)\}$,
$S_x=\{\sgrad \phi_a(x)\} \cap \{\sgrad I_i(x)\}$ and
the foliation $\mathcal L\vert_{\OO_1}$ does not depend on $\varphi$).

Let $\pi: \Phi^{-1}(0)\to \Phi^{-1}(0)/G=T^*(Q/G)$ be the natural projection
and let
$\tilde \UU=\pi(\UU_1)\subset T^*(Q/G)$.
The foliation $\mathcal L$ induces a foliation
$\tilde{\mathcal T}=\pi(\mathcal L)$ of $\tilde \UU$
by an $(r-s)$--dimensional invariant  manifolds of the geodesic flow of the
submersion metric.
We shall see below that
$\tilde{\mathcal T}$ is a foliation of $\tilde \UU$ by an invariant
tori with respect to certain complete algebra of first
integrals $\FF_0$.

The foliation  $\mathcal L$ can be seen as the level sets of
of $G$--invariant integrals $f_1,\dots,f_\rho$ on $\UU_1$,
$\rho=\dim T^*Q-\dim G+s-r$.
Indeed, this is always true locally:
for $\sigma$ small enough there are functions
$f_i=f_i(I,p,q)$ on $\OO_1$
such that the tangent spaces $T_x\mathcal L$
(recall that $\mathcal L\vert_{\OO_1}$
does not depend on $\varphi$)
are given by the equations
$$
T_x\mathcal L=\{\xi\in T_x \OO_1, \; df_i(x)(\xi)=0, \; i=1,\dots,\rho\}.
$$
Then extend the $f_i$ to $G$--invariant functions on $\UU_1$.
Let $\FF_0$ be the induced algebra of first integrals in $\tilde\UU$.
Since
$$
D_{\pi(x)} =\{\xi\in T_{\pi(x)} \tilde U,
\; df_0(\pi(x))(\xi)=0,\; f_0\in\FF_0\} =
T_{\pi(x)}\tilde{\mathcal T}=d\pi(x)(T_x\mathcal L),
$$
$x\in \UU_1$ are isotropic spaces,
by remark 1 we get that $\FF_0$ is complete in $\tilde \UU$.
Note that
\begin{equation}
\ddim\FF_0=\dim T^*(Q/G)+s-r,\quad \dind\FF_0=r-s.
\label{**}
\end{equation}

Now  fill up $T^*(Q/G)$ with
countably many
disjoint toroidal domains
$\tilde{\OO}_\alpha=\T^{r(\alpha)}\{\tilde\varphi\}
\times B_{\sigma(\alpha)}\{\tilde I,\tilde p,\tilde q\}$
(with possible different dimensions of tori).
In every domain $\tilde{\OO}_\alpha$,
one can construct complete involutive set of integrals that can be then
"glued" in order to obtain a complete involutive set of
integrals globally defined.
The construction is suggested by Brailov for Darboux symplectic balls.
We follow the construction in \cite{BJ}.

We have the following $m=\dim Q/G$ commuting integrals in $\tilde{\OO}_\alpha$:
$$
h_1=\tilde{I}_1^2,\dots,h_r=\tilde{I}_r^2, \;
h_{r+1}=\tilde{p}_1^2+\tilde{q}_1^2,\dots,h_m=\tilde{p}_k^2+\tilde{q}_k^2.
$$

Let $g_\alpha: \Bbb{R}\to\Bbb{R}$ be a smooth nonnegative function
such that $g_\alpha(x)$ is equal to zero for $\vert x \vert > \sigma(\alpha)$,
$g_\alpha$ monotonically increases on $[-\sigma(\alpha),0]$
and monotonically decreases on $[0,\sigma(\alpha)]$.
Let $h_\alpha(y)=g_\alpha(h_1(y)+\dots+h_m(y))$.
This function can be extended
by zero to the whole manifold $T^*(Q/G)$.
Then $f_i^\alpha=h_\alpha\cdot h_i$, $i=1,\dots,n$
will be commuting functions,
independent on an open dense subset of $\tilde{\OO}_\alpha$.
With a "good" choice of $g_\alpha$'s,
a complete commutative set of smooth integrals
 is given by
$f_i(y)=f_i^\alpha(y)$ for $y \in \tilde{\OO}_\alpha\subset \cup_\beta \tilde{\OO}_\beta$
and zero otherwise, $i=1,\dots,m$.
\end{proof}

\begin{remark}
It is clear that
a similar statement holds for an arbitrary
Hamiltonian $G$--space $(M,\omega)$ ($G$ acts freely on $M$) and
an integrable Hamiltonian system $\dot x=\sgrad H(x)$
with compact iso-energy levels $M_h=H^{-1}(h)$.
The $M$ is foliated by invariant tori in the open dense
set $\reg M$. If $\reg M$ intersects the submanifold
$\Phi^{-1}(\eta)$ in a dense set then
the reduced Hamiltonian system on $(M_\eta,\omega_\eta)$
will be completely integrable.
\end{remark}

\smallskip

Here is a simple construction that gives examples satisfying
the hypotheses of theorem 4.
Suppose we are given Hamiltonian $G$--actions
on two symplectic manifolds $(M_1,\omega_1)$ and $(M_2,\omega_2)$
with moment maps $\Phi_{M_1}$ and $\Phi_{M_2}$.
Then we have the natural diagonal action of $G$ on the product
$(M_1\times M_2,\omega_1\oplus\omega_2)$, with moment map
\begin{equation}
\Phi_{M_1\times M_2}=\Phi_{M_1}+\Phi_{M_2}.
\label{3.2}
\end{equation}

If $(Q_1,g_1)$ and $(Q_2,g_2)$ have integrable geodesic flows, then
$(Q_1\times Q_2, g_1\oplus g_2)$ also has
integrable geodesic flow
and   $\reg T^*(Q_1\times Q_2)=\reg T^*Q_1 \times \reg T^*Q_2$.
Using (\ref{3.2}),
one can easily see that if the $G$--actions
on $Q_1$ and $Q_2$ are almost everywhere locally free,
then  a generic horizontal vector of the submersion
$Q_1 \times Q_2 \to Q_1 \times_G Q_2= (Q_1\times Q_2)/G$
belongs to $\reg T^*(Q_1 \times Q_2)$.
Thus we get the following statement.

\begin{cor}
Let the Lie group $G$ act by isometries
on the compact Riemannian manifolds
$(Q_1,g_1)$ and $(Q_2,g_2)$ almost everywhere locally freely,
and freely on the product $Q_1 \times Q_2$.
Suppose that the geodesic flows on $Q_1$ and $Q_2$
are completely integrable.
Then the geodesic flow on $Q_1 \times_G Q_2$,
endowed with the submersion metric, will be completely integrable.
\end{cor}

The theorem and corollary allow us to construct examples of manifolds
with completely integrable geodesic flows starting from some known
integrable cases.
This generalizes the
construction of Paternain and Spatzier of
manifolds of the form $Q\times_{S^1} N$, where $N$ is a surface of revolution
and the geodesic flow on $Q$ possesses a
complete involutive algebra of $S^1$--invariant integrals \cite{PS}.

\begin{example}
Suppose the Lie group $G$ acts freely by
isometries on $(Q,g)$.
Let $G_1$ be an arbitrary compact Lie group, which contains
$G$ as a subgroup. Let $ds^2_1$ be some
left-invariant Riemannian metric on $G_1$
with integrable geodesic flow (see example 10 below). Then $G$ acts in the natural way by isometries on $(G_1,ds^2_1)$.
Therefore, by corollary 6, if the geodesic flow on $Q$ is completely integrable,
then the geodesic flows on $Q\times_G Q$
and $Q \times_G G_1$ endowed with the submersion metrics
will be also completely integrable.
\end{example}

As an illustration,
we will show that we can get
a very simple proof
for the complete integrability of the geodesic flows on
a class of bi-quotients of Lie groups.

\section{Bi-quotients of compact Lie groups}

Let $G$ be a compact connected Lie group and $\mathfrak g$ be
its Lie algebra.
Let $\langle \cdot,\cdot \rangle$ be an
$Ad_G$--invariant
scalar product on  $\g$ and
$ds^2_0$ the corresponding bi-invariant metric on $G$.
In what follows we shall identify $T^*G$ and $TG$ by the bi-invariant metric.

Consider a connected subgroup $U$ of $G\times G$
and define the action of $U$ on $G$ by:
\begin{equation}
(g_1,g_2)\cdot g =g_1gg_2^{-1}, \quad (g_1,g_2)\in U,\; g\in G.
\label{***}
\end{equation}

If the action is free then the orbit space $G/U$ is a
smooth manifold called a
{\it bi-quotient} of the Lie group $G$.
In particular, if $U=K\times H$, where
$K$ and $H$ are subgroups of $G$, then
the bi-quotient $G/U$
is denoted by $K\backslash G/H$.
The bi-invariant metric $ds^2_0$ on $G$, induces the
submersion metric $ds^2_{0,sub}$ on $G/U$.
In this way, via the submersion $Sp(2)\to\Sigma^7$,
Gromoll and Meyer \cite{GM} obtained
an exotic 7--sphere $\Sigma^7$.

It is well known that the geodesic flow of the metric $ds^2_0$ on $G$
is completely integrable in the non-commutative sense.
For a complete algebra of integrals we can take
$\FF=\FF_1+\FF_2$, where $\FF_1$ and $\FF_2$
are the functions obtained from polynomials on $\g$ by
left and right translations respectively.
Then $\ddim \FF=2\dim G-\rank G$, $\dind \FF=\rank G=r$.
The system is $\T^r$--dense completely integrable.
Let  $\mathbb{R}[\mathfrak g]^G$ be the algebra
of the invariant
polynomials on $\mathfrak g$.
The algebra  $\mathbb{R}[\mathfrak g]^G$ is generated by
$r$ homogeneous polynomials $p_1,\dots,p_{r}$.
Let $f_1,\dots,f_r$ be the left (or right) translations of
$p_1,\dots,p_{r}$ to $TG$.
Then $f_1,\dots,f_r$ generate $\FF_1\cap \FF_2$, and can be seen
as "action" variables:
the vector fields $X_1=\sgrad f_1,\dots,X_r=\sgrad f_r$ form a basis
of commuting vector fields on the regular invariant tori.

Let $\mathfrak u\subset \g \cong T_e G$
be the  vertical space at the neutral element of the group.
Then the horizontal space $\mathfrak v$ at the neutral element
is the orthogonal complement of $\mathfrak u$ with respect to $\langle\cdot,\cdot\rangle$.

Let $\g_\xi=\{ \eta\in\g,\; [\xi,\eta]=0\}$.
The element $\xi\in \g$ is called {\it regular} if the
adjoint orbit $O_G(\xi)$ of the $G$--action has a maximal dimension
(equal to $\dim G-\rank G$),
or equivalently if $\g_\xi$ is an $r$--dimensional commutative
subalgebra of $\g$.
Then
$\g_\xi=\{ \nabla p(\xi), \; p\in \mathbb{R}[\mathfrak g]^G\}$.
The element $\xi\in\g$ is called {\it singular} if it is not regular.

\begin{thm}
Suppose that the horizontal space $\mathfrak v\subset\g$
contains a regular element of the Lie algebra $\g$.
Then the geodesic flow of the metric $ds^2_{0,sub}$ on
the  bi-quotient $G/U$ is completely integrable.
Moreover,
the phase space $T(G/U)$
is almost everywhere foliated by
invariant isotropic tori of dimension
$$
\rank G - \min_{\xi\in \mathfrak v} \dim \mathfrak u_\xi,
$$
where $\mathfrak u_{\xi}=
\{\eta\in \mathfrak u, \; [\eta,\xi]=0\}\subset \mathfrak g_\xi$.
\end{thm}

\begin{proof}
Since $\FF$ consists of analytic functions, if one horizontal
vector belongs to $\reg TG$, then this property will hold
for a general horizontal vector as well.
Furthermore,  it is clear that the regular elements of $\g$
belong to $\reg TG$.
Therefore, from theorem 4 we obtain the integrability of the geodesic flow.

To complete the proof, we have to find the dimension of the space (\ref{*}).
Let $\xi\in \mathfrak v\subset \mathfrak g \cong T_e G$ be generic.
Then the tangent space $T_\xi \T^r$ to the invariant torus containing $\xi$
can be naturally identified with $\mathfrak g_\xi$ and
the space $S_\xi=T_\xi(U\cdot \xi) \cap T_\xi \T^r$
can be naturally identified with $\mathfrak u_\xi$.
By (\ref{**}), it follows that the $r$--dimensional invariant tori are reduced
to $(r-\dim \mathfrak u_\xi)$--dimensional invariant tori.
\end{proof}

Without using submersions, the complete integrability
of the geodesic flows of metrics $ds^2_{0,sub}$
on homogeneous spaces $G/H$ and
bi-quotients $K\backslash G/H$ has been proved
in \cite{BJ1} and \cite{BJ}, respectively.

Theorem 8 includes the
examples given by Paternain and  Spatzier \cite{PS}
and Bazaikin \cite{Baz}.
Indeed, for the submersion examples studied in \cite{Baz},
Bazaikin already proved the regularity of generic $\xi\in\mathfrak v$.
Paternain and  Spatzier proved the integrability of the geodesic flows
for the  Eschenburg examples $M^7_{1,-1,2m,2m}$ and for
the Gromoll and Meyer 7--sphere $\Sigma^7$.
Since the set of singular elements has codimension at least 3 in $\g$,
the condition of theorem 7 is automatically satisfied for the
Eschenburg examples  $SU(3)\to M^7_{k,l,p,q}$.
Here $U=U_{k,l,p,q}\cong T^1 \subset T^2\times T^2$, where
$T^2$ is a maximal torus (see \cite{E}).
On the other side,
one can easily see that independence of the integrals $f_4$ and $f_5$
for the geodesic flow on $\Sigma^7$ (page 361, \cite{PS})
gives us independence of the invariant polynomials
$tr(\xi^2)$ and $tr(\xi^4)$ at generic $\xi\in\mathfrak v$.
Thus a generic $\xi\in\mathfrak v$ is a regular element of $\mathfrak{sp}(2)$.

The general construction presented here leads to
smooth commuting integrals while the commuting integrals
given in \cite{PS, Baz}
are analytic functions.
On the other hand, we proved that the Lagrangian
tori are fibered into invariant isotropic tori,
so in this sense the system is degenerate.
For example,
the 14--dimensional manifolds $TM^7_{k,l,p,q}$ are foliated by
two-dimensional tori.

\begin{remark}
The exotic 7--sphere $(\Sigma^7,ds^2_{0,sub})$ admits
an effective action of $O(2)\times SO(3)$ by isometries \cite{GM}.
The existence of the non-abelian group of isometries is
related to the {non-commutative} integrability of the geodesic flow.
Namely, suppose that a compact connected Lie group $G$ acts effectively
by isometries on the
$n$--dimensional Riemannian manifold $(Q,g)$
with moment map $\Phi: T^*Q\to\g^*$.
If the geodesic flow is $\T^n$--dense completely integrable,
then in every toroidal domain the functions
$\phi_a(x)$, $a\in\g$
depend only on the action variables and so $\{\phi_a,\phi_b\}=\phi_{[a,b]}$
vanishes on $\reg T^*Q$.
Therefore $[\g,\g]=0$ and the Lie group $G$ is a torus.
\end{remark}

\smallskip

In order to obtain manifolds with strictly positive sectional
curvature, Eschenburg considered a one--parameter family of
left-invariant metrics $ds^2_t$ on $SU(3)$ which are different
from the bi-invariant metric $ds^2_0$  \cite{E}. One can prove
that the geodesic flows of the metrics $ds^2_t$ are completely
integrable and that we can apply theorem 4 to get the
integrability of the geodesic flows of the submersion metrics
$ds^2_{t,sub}$ on $M^7_{k,l,p,q}$. A similar multi-dimensional
example of bi-quotients with metrics different from $ds^2_{0,sub}$
and integrable geodesic flows is presented below.

\begin{example} \footnote{The example is slightly different
than example 10 in the journal version.} There are several
constructions of left(right)-invariant metrics on Lie groups with
integrable geodesic flows (see \cite{TF}). We shall use the
following construction due to Mishchenko and Fomenko. Let $G$ be a
compact connected Lie group, $T\subset G$ a maximal torus, and
$\g$ and $\mathfrak{t}$ the corresponding Lie algebras. Take
$a_1,a_2$, $b_1,b_2\in\mathfrak t$ such that $a_1$ and $a_2$ are
regular elements of $\g$, i.e, $\g_{a_1}=\g_{a_2}=\mathfrak t$.
Let $D_1, D_2: \mathfrak t\to \mathfrak t$ be symmetric operators.
Denote by $\varphi_1$ and $\varphi_2$ the symmetric operators
(called {\it sectional operators} \cite{TF}) defined according to
the orthogonal decomposition: $\mathfrak g=\mathfrak t+\mathfrak
t^{\bot}$:
$$
\varphi_{i}\vert_{\mathfrak t}=D_i,
\quad \varphi_{i}\vert_{\mathfrak t^{\bot}}=ad_{a_i}^{-1}ad_{b_i}, \quad i=1,2.
$$
In the case of compact Lie groups, among the sectional operators there are
positive definite ones. Then
the Hamiltonian functions $H_1$ and $H_2$,
obtained from quadratic forms
$$
B_i(\xi,\xi)=\langle \varphi_{i} \xi,\xi\rangle, \quad \xi\in\g,
\quad i=1,2
$$
by left and right translations, define left-invariant and
right-invariant metrics on $G$ which we shall denote by $ds_{1}^2$
and $ds^2_2$, respectively. Mishchenko and Fomenko proved that the
geodesic flows of $ds_{1}^2$ and $ds^2_2$ are completely
integrable \cite{TF}. But now we can take the sum $H_1+H_2$ which
also gives the metric $ds^2$ on $G$ with completely integrable
geodesic flow \cite{BJ1}. Let $U$ be any subgroup of $T\times T$
such that the action (\ref{***}) is free. It can be proved that
$U$ acts on $(G,ds^2)$ by isometries and that a generic horizontal
vector of the submersion $G \to G/U$ belongs to $\reg TG$. Thus,
by theorem 4, the geodesic flow of the metric $ds^2_{sub}$ on the
bi-quotient $G/U$ is completely integrable. The motion of the
system in $T(G/U)$ is more complicated than in the case of the
metric $ds^2_{0,sub}$ and proceeds along tori that have the
dimension of $G/U$ in the general case. In particular, when
$U=\{e\}\times T$, $ds^2_{sub}$ is a Riemannian metric on the flag
manifold $G/T$. The integrability of the geodesic flow of this
metric, was proved in a different way in \cite{BJ1} (theorem 5).
\end{example}

\subsection*{Acknowledgments}
I am grateful to Prof. A. V. Bolsinov for the
very useful remarks and discussions which helped me look at the problem
from a general point of view.
This letter was written during my stay at
the Mathematisches Institut LMU, M\" unchen, as a
postdoc of the
Graduiertenkolleg "Mathematik im Bereich ihrer Wechselwirkung mit der Physik".
I would like to thank
Ludwig--Maximillians--Universit\"at for the hospitality,
Prof. K. Cieliebak for the kind support and the referee
for various useful suggestions which improved the exposition
of the paper.
The research was partially supported by the Serbian Ministry
of Science and Technology, Project 1643 -- Geometry and Topology
of Manifolds and Integrable Dynamical Systems.

\smallskip

\noindent {\bf Note Added.}
During the refereeing process
of this letter, N. T. Zung's preprint \cite{Zung}
appeared which contains a result somewhat more general
than remark 5
(the manifold $M/G$ is allowed to have singularities),
obtained independently.

\bibliographystyle{amsplain}

\end{document}